\documentclass[draft]{elsart}
\usepackage{amssymb}
\usepackage{graphicx}

\input{tcilatex}

\begin{document}

\begin{frontmatter}
\title{Pressure control system with nonlocal friction}
\author[kwok]{Kwok Sau Fa}, 
\ead{kwok@dfi.uem.br}
\author[sara]{S. R. Osipi}

\address[kwok]{Departamento de F\'{\i}sica, Universidade Estadual de Maring\'{a}, Av.
Colombo 5790, 87020-900, \ Maring\'{a}-PR, Brazil, Tel.: 55 44 32614330}

\address[sara]{Departamento de Engenharia Química, Universidade Estadual de Maring\'{a}, Av.
Colombo 5790, 87020-900, \ Maring\'{a}-PR, Brazil}

\begin{abstract}
We analyze the motion of a pressure control system  described by a differential
equation with nonlocal dissipative force. This system is composed by an oscillator,
 a membrane and a constant force. We consider the dissipative memory kernel consisting of
two terms. One of them is described by the Dirac delta function which
represents a local friction, whereas for the second one we consider two
types: the exponential and power-law functions which represent nonlocal
dissipative forces. For these cases, one can obtain exact solutions for the
displacement and velocity. The long-time behaviors of these quantities
are also investigated. 
\end{abstract}

\end{frontmatter}

\section{Introduction}

Vibration is an ubiquitous phenomenon in nature and one can perceive it in
many macroscopic systems by the senses. In the microscopic scale this
phenomenon is also present, for instance, the vibration of atoms inside of
each molecule or vibration of molecules around their equilibrium positions
in \ a solid body. In engineering the vibration is present in structural and
flexible multibody systems. Usually these systems are accompanied by
dissipation process. General speaking the dissipation process may be local
or nonlocal. The nonlocal friction is related to memory effect which depends
on the past motion of a system. This kind of friction has been used to study
anomalous diffusion processes which have been observed in various kinds of
systems such as bacterial cytoplasm motion \cite{cox}, conformational
fluctuations within a single protein molecule \cite{kou,min} and
fluorescence intermittency in single enzymes \cite{chaudhury}. Further, it
has been applied to structural damping systems, viscoelastic materials,
seismic and vibration isolation (see \cite%
{bagley,makris,shiti1,shiti2,mainardi,agrawal,adhikari} and references
therein). In fact, nonlocal friction may be used to enhance the modelling of
damping systems.

In this work we consider a system of vibration composed by an oscillator, a
membrane and a constant force. This simple system can be used as a pressure
control device such as shown in Fig. 1 \cite{kreys}. We note that the
pressure may be exerted by a fluid, some mechanical mechanism or a constant
force such as the gravitational force of the earth. In this last case the
motion of the system should be positioned, for convenience, in the same
direction of the gravitational force. However, our purpose is to employ this
system to investigating the dissipation process by the action of membrane in
the system. For elastic membrane one can consider the system
membrane-oscillator as an oscillator. If the system presents nonlocal
dissipation due to membrane then the mass coupled to the oscillator will
describe different motion in comparison with that system with local
dissipation.

In order to model the above system we consider the following differential
equation:

\begin{equation}
m\frac{\text{d}v}{\text{d}t}+\int_{0}^{t}dt_{1}\gamma \left( t-t_{1}\right)
v(t_{1})+kx=p\theta (t)\text{ , }  \label{eq1}
\end{equation}%
where $m$ is the effective mass, $k$ is the spring constant, $v$ is the
velocity,\ $\gamma (t)$ is the dissipative memory kernel and $\theta (t)$ is
the Heaviside function. We note that the dissipative memory kernel $\gamma
(t)$ may be a very complicated function, but we will restrict it for some
particular cases. We will consider that the friction force is described by a
combination of local and nonlocal forces. The dissipative memory kernel for
a local friction is represented by the Dirac $\delta $ function, whereas for
the nonlocal friction we consider the exponential and power-law functions.
For this linear system, general expression for the solution can be obtained.
We will present the explicit solutions for the relaxation function. We will
also analyze the long-time behavior of the solutions.

\bigskip

\section{Exact solutions and analysis of results}

Before analyzing the details of the above system for some specific
dissipative memory kernel, we present general expression for the solution of
Eq. (\ref{eq1}). Eq. (\ref{eq1}) is linear then it can be solved by using
the Laplace transform, with the initial conditions $x_{0}=x(0)$ and $%
v_{0}=v(0)$. The application of the Laplace transform to Eq. (\ref{eq1})
yields

\begin{equation}
\overline{x}(s)=\frac{x_{0}\left( ms+\overline{\gamma }\right) +mv_{0}+p/s}{%
ms^{2}+s\overline{\gamma }+k}\text{ ,}  \label{eq4}
\end{equation}%
where $\overline{\gamma }$ is the Laplace transform of the damping kernel $%
\gamma (t)$. Now we rewrite Eq. (\ref{eq4}) as follows:

\begin{equation}
\overline{x}(s)=\left( x_{0}\left( ms+\overline{\gamma }\right)
+mv_{0}+p/s\right) \overline{G}(s)  \label{eq5}
\end{equation}%
where $G(t)$ is the relaxation function and it is the Laplace inversion of

\begin{equation}
\overline{G}(s)=\frac{1}{ms^{2}+s\overline{\gamma }+k}\text{ .}  \label{eq6}
\end{equation}%
For simplicity, one considers $x_{0}=v_{0}=0$, then Eq. (\ref{eq5}) becomes

\begin{equation}
\overline{x}(s)=\frac{p}{s}\overline{G}(s)\text{ .}  \label{eq7}
\end{equation}%
Moreover, from Eq.(\ref{eq7}) one can obtain $\overline{v}(s)$ which is
given by

\begin{equation}
\overline{v}(s)=p\overline{G}(s)\text{ .}  \label{eq8}
\end{equation}%
From this last equation we arrive at

\begin{equation}
v(t)=pG(t)  \label{eq9}
\end{equation}%
and

\begin{equation}
x(t)=p\int_{0}^{t}dt_{1}G\left( t_{1}\right) \text{ .}  \label{eq10}
\end{equation}%
We see that the velocity is proportional to the relaxation function $G(t)$.
In order to obtain explicit solutions for $x(t)$ and $v(t)$ we should give
the form of dissipative memory kernel $\gamma (t)$. We consider three cases.

\textit{First case.} We take the dissipation process as local friction. In
this case the dissipative memory kernel is represented by the Dirac $\delta $
function and the friction force is proportional to $v(t)$. The function $%
\gamma $ is given by $\gamma (t)=\gamma _{\delta }\delta (t)$ and its
Laplace transform is $\overline{\gamma }(s)=\gamma _{\delta }$. Substituting
it into Eq. (\ref{eq6}) we obtain \cite{kreys}

\begin{equation}
x(t)=\frac{p}{k}\left\{ 1-\left[ \cos \left( \omega _{1}t\right) +\frac{%
\alpha }{\omega _{1}}\sin \left( \omega _{1}t\right) \right] \exp \left(
-\alpha t\right) \right\} \text{ ,\ \ }\gamma _{\delta }^{2}<4mk\text{ ,\ }
\label{eq10a}
\end{equation}%
\begin{equation}
v(t)=pG(t)=\frac{p}{m\omega _{1}}\sin \left( \omega _{1}t\right) \exp \left(
-\alpha t\right) \text{ , \ }\gamma _{\delta }^{2}<4mk\text{ ,}
\label{eq10b}
\end{equation}%
\begin{equation}
x(t)=\frac{p}{k}\left\{ 1-\left( 1+\alpha t\right) \exp \left( -\alpha
t\right) \right\} \text{ , \ \ }\gamma _{\delta }^{2}=4mk\text{ ,}
\label{eq10c}
\end{equation}%
\begin{equation}
v(t)=pG(t)=\frac{p}{m}t\exp \left( -\alpha t\right) \text{ , \ }\gamma
_{\delta }^{2}=4mk\text{ ,}  \label{eq10d}
\end{equation}%
\begin{equation}
x(t)=\frac{p}{k}\left\{ 1-\left[ \cosh \left( \omega _{2}t\right) +\frac{%
\alpha }{\omega _{2}}\sinh \left( \omega _{2}t\right) \right] \exp \left(
-\alpha t\right) \right\} \text{ , \ }\gamma _{\delta }^{2}>4mk\text{ ,}
\label{eq10e}
\end{equation}%
\begin{equation}
v(t)=\frac{p}{m\omega _{2}}\sinh \left( \omega _{2}t\right) \exp \left(
-\alpha t\right) \text{, \ }\gamma _{\delta }^{2}>4mk\text{ ,}  \label{eq10f}
\end{equation}%
where $\alpha =\gamma _{\delta }/2m$, $\omega _{1}=\sqrt{\frac{k}{m}-\alpha
^{2}}$ and $\omega _{2}=\sqrt{\alpha ^{2}-\frac{k}{m}}$. The solutions
present three damped cases: Underdamped $\gamma _{\delta }^{2}<4mk$,
critical damping $\gamma _{\delta }^{2}=4mk$ \ and overdamped $\gamma
_{\delta }^{2}>4mk$. All the three cases have the same final equilibrium
position which is given by $x_{f}=p/k$. Their behaviors are shown in Figs. 2
and 3. We see that the solution (\ref{eq10c}) decays faster than other
cases. This means that the oscillator with a critical damping moves to its
final equilibrium position faster than other cases. Further, for the
underdamped and critical damping cases the velocity reaches a maximum value
and then it decreases to zero continuously.

\textit{Second case}. We take the combination of a local friction
represented by the Dirac $\delta $ function and exponential function for the
\ frictional memory kernel which is given by $\gamma (t)=\gamma _{\delta
}\delta (t)+\gamma _{\lambda }e^{-\lambda t}$. We note that the exponential
damping model has been used to model \ linear elastic and viscoelastic
systems (see \cite{adhikari} and references therein). The Laplace transform
of $\gamma (t)$ is $\overline{\gamma }(s)=\gamma _{\delta }+\gamma _{\lambda
}/\left( s+\lambda \right) $. Then we obtain from Eqs. (\ref{eq6}), (\ref%
{eq9}) and (\ref{eq10}) the following results:

\[
x(t)=\frac{pt^{2}}{m}\sum_{n=0}^{\infty }\frac{\left( -\frac{k\lambda }{m}%
t^{3}\right) ^{n}}{n!}\sum_{j=0}^{\infty }\frac{\left( -\frac{\left( \gamma
_{\delta }\lambda +\gamma _{\lambda }+k\right) }{m}t^{2}\right) ^{j}}{j!} 
\]%
\begin{equation}
\times \left[ E_{1,3+j+2n}^{(n+j)}\left( -\left( \lambda +\frac{\gamma
_{\delta }}{m}\right) t\right) +\lambda tE_{1,4+j+2n}^{(n+j)}\left( -\left(
\lambda +\frac{\gamma _{\delta }}{m}\right) t\right) \right] \text{ }
\label{eq11}
\end{equation}%
and%
\[
v(t)=\frac{pt}{m}\sum_{n=0}^{\infty }\frac{\left( -\frac{k\lambda }{m}%
t^{3}\right) ^{n}}{n!}\sum_{j=0}^{\infty }\frac{\left( -\frac{\left( \gamma
_{\delta }\lambda +\gamma _{\lambda }+k\right) }{m}t^{2}\right) ^{j}}{j!} 
\]%
\begin{equation}
\times \left[ E_{1,2+j+2n}^{(n+j)}\left( -\left( \lambda +\frac{\gamma
_{\delta }}{m}\right) t\right) +\lambda tE_{1,3+j+2n}^{(n+j)}\left( -\left(
\lambda +\frac{\gamma _{\delta }}{m}\right) t\right) \right] \text{ ,}
\label{eq11a}
\end{equation}%
where $E_{\beta ,\delta }\left( y\right) $ is the generalized Mittag-Leffler
function \cite{gorenflo,erdel} defined by

\begin{equation}
E_{\beta ,\delta }\left( y\right) =\sum_{n=0}^{\infty }\frac{y^{n}}{\Gamma
\left( \beta n+\delta \right) }\text{, }\beta >0\text{, }\delta >0\text{ ,}
\label{eq12}
\end{equation}

\begin{equation}
E_{\beta ,\delta }^{(k)}\left( y\right) =\frac{\text{d}^{k}}{\text{d}y^{k}}%
E_{\beta ,\delta }\left( y\right) =\sum_{n=0}^{\infty }\frac{(n+k)!y^{n}}{%
n!\Gamma \left( \beta (n+k)+\delta \right) }\text{ .}  \label{eq13}
\end{equation}%
and $\Gamma \left( z\right) $ is the Gamma function.

Note that relevant information of these quantities can be obtained by
analyzing their asymptotic behaviors. Therefore we use the long-time limit
of the generalized Mittag-Leffler function \cite{erdel} given by

\begin{equation}
E_{\alpha ,\beta }(z)\sim -\frac{1}{z\Gamma \left( \beta -\alpha \right) }%
\text{ .}  \label{eq16}
\end{equation}%
From Eqs. (\ref{eq11}), (\ref{eq11a}) and (\ref{eq16}) yield

\begin{equation}
x(t)\sim \frac{p}{k}\left[ 1-\frac{\gamma _{\delta }\lambda +\gamma
_{\lambda }}{\gamma _{\delta }\lambda +\gamma _{\lambda }+k}\exp \left( -%
\frac{k\lambda t}{\gamma _{\delta }\lambda +\gamma _{\lambda }+k}\right) %
\right] \text{ ,}  \label{eq17a}
\end{equation}%
and%
\begin{equation}
v(t)\sim \frac{p\lambda \left( \gamma _{\delta }\lambda +\gamma _{\lambda
}\right) }{\left( \gamma _{\delta }\lambda +\gamma _{\lambda }+k\right) ^{2}}%
\exp \left( -\frac{k\lambda t}{\gamma _{\delta }\lambda +\gamma _{\lambda }+k%
}\right) \text{ ,}  \label{eq17b}
\end{equation}%
where we consider the fact that $\left( \lambda +\gamma _{\delta }/m\right)
t\gg 1$ and $(\lambda \gamma _{\delta }+\gamma _{\lambda }+k)\gtrsim m\left(
\lambda +\gamma _{\delta }/m\right) ^{2}$. We see that the displacement $%
x(t) $\ and velocity $v(t)$ decay exponentially similar to those obtained
from the system with an instantaneous friction. Moreover, we note that the
constant force does not modify the exponential asymptotic form. From Eq. (%
\ref{eq17a}) we can obtain the final equilibrium position which is given by $%
x_{f}=p/k$. In Fig. 4 we verify the asymptotic result of $x(t)$ Eq. (\ref%
{eq11}) by comparing with Eq. (\ref{eq17a}), whereas in Fig. 5 we verify the
asymptotic result of $v(t)$ Eq. (\ref{eq11a}) by comparing with Eq. (\ref%
{eq17b}). The asymptotic results are very close to the exact ones.

\textit{Third case}. We consider the combination of instantaneous friction
represented by the Dirac $\delta $ function and long-time frictional memory
kernel given by $\gamma (t)=\gamma _{\delta }\delta (t)+\gamma _{\beta
}t^{-\beta },$ for $0<\beta <1$. Then, the Laplace transform of $\gamma (t)$
is $\overline{\gamma }=\gamma _{\delta }+\gamma _{\beta }^{\ast }s^{\beta
-1} $, where $\gamma _{\beta }^{\ast }=\gamma _{\beta }\Gamma (1-\beta )$ .
We note that the second term of $\gamma (t)$ describes a fractional
derivative on Eq. (\ref{eq1}) which corresponds to the Caputo fractional
derivative \cite{gorenflo}. The solutions for the displacement and velocity
can be obtained from Eqs. (\ref{eq6}), (\ref{eq9}) and (\ref{eq10}). The
results are given by

\begin{equation}
x(t)=\frac{pt^{2}}{m}\sum_{n=0}^{\infty }\frac{\left( -\frac{k}{m}%
t^{2}\right) ^{n}}{n!}\sum_{j=0}^{\infty }\frac{\left( -\frac{\gamma _{\beta
}^{\ast }}{m}t^{2-\beta }\right) ^{j}}{j!}E_{1,3+n+(1-\beta
)j}^{(n+j)}\left( -\frac{\gamma _{\delta }}{m}t\right) \text{ }  \label{eq20}
\end{equation}%
and

\begin{equation}
v(t)=\frac{pt}{m}\sum_{n=0}^{\infty }\frac{\left( -\frac{k}{m}t^{2}\right)
^{n}}{n!}\sum_{j=0}^{\infty }\frac{\left( -\frac{\gamma _{\beta }^{\ast }}{m}%
t^{2-\beta }\right) ^{j}}{j!}E_{1,2+n+(1-\beta )j}^{(n+j)}\left( -\frac{%
\gamma _{\delta }}{m}t\right) \text{ .}  \label{eq20a}
\end{equation}%
For $\gamma _{\delta }=0$ (nonlocal friction) we have

\begin{equation}
x(t)=\frac{pt^{2}}{m}\sum_{n=0}^{\infty }\frac{\left( -\frac{k}{m}%
t^{2}\right) ^{n}}{n!}E_{2-\beta ,3+\beta n}^{(n)}\left( -\frac{\gamma
_{\beta }^{\ast }}{m}t^{2-\beta }\right)  \label{eq20b}
\end{equation}%
and

\begin{equation}
v(t)=\frac{pt}{m}\sum_{n=0}^{\infty }\frac{\left( -\frac{k}{m}t^{2}\right)
^{n}}{n!}E_{2-\beta ,2+\beta n}^{(n)}\left( -\frac{\gamma _{\beta }^{\ast }}{%
m}t^{2-\alpha }\right) \text{ .}  \label{eq20c}
\end{equation}%
On the other hand, for $\gamma _{\beta }=0$ (local friction) we have%
\begin{equation}
x(t)=\frac{pt^{2}}{m}\sum_{n=0}^{\infty }\frac{\left( -\frac{k}{m}%
t^{2}\right) ^{n}}{n!}E_{1,3+n}^{(n)}\left( -\frac{\gamma _{\delta }}{m}%
t\right)  \label{eq20d}
\end{equation}%
and%
\begin{equation}
v(t)=\frac{pt}{m}\sum_{n=0}^{\infty }\frac{\left( -\frac{k}{m}t^{2}\right)
^{n}}{n!}E_{1,2+n}^{(n)}\left( -\frac{\gamma _{\delta }}{m}t\right) \text{ .}
\label{eq20e}
\end{equation}%
We have numerically compared these last solutions (\ref{eq20d}) and (\ref%
{eq20e}) with (\ref{eq10a})-(\ref{eq10f}) and they give similar results.
This means that the solutions (\ref{eq20d}) and (\ref{eq20e}) contain the
underdamped, critical damping and overdamped solutions.

Now we analyze the asymptotic behavior of the above quantities $x(t)$ and $%
v(t)$. By using the long-time limit of $E_{\alpha ,\beta }(y)$ (\ref{eq16})
we obtain

\begin{equation}
x(t)\sim \frac{p}{k}\left[ 1-\frac{\gamma _{\beta }^{\ast }\sin \left( \pi
\beta \right) \Gamma \left( \beta \right) }{k\pi t^{\beta }}\right] \text{ ,}
\label{eq21a}
\end{equation}

\begin{equation}
v(t)\sim \frac{p\gamma _{\beta }^{\ast }\sin \left( \pi \beta \right) \Gamma
\left( 1+\beta \right) }{k^{2}\pi t^{1+\beta }}\text{ ,}  \label{eq21b}
\end{equation}%
where we consider the fact that $\gamma _{\delta }t/m\gg 1$, $\gamma _{\beta
}^{\ast }m^{(1-\beta )}\gtrsim \gamma _{\delta }^{(2-\beta )}$ and $%
km^{\beta }\gtrsim \gamma _{\beta }^{\ast }\gamma _{\delta }^{\beta }$. We
can see that these asymptotic results decay as a power-law and they are
dominated by the nonlocal dissipative force; The parameter $\gamma _{\delta
} $ does not appears in these leading terms. From Eq. (\ref{eq21a}) we can
obtain the final equilibrium position which is given by $x_{f}=p/k$. This
means that the nonlocal dissipative force described by a long-time memory
kernel may suppress the presence of a local friction in the long-time limit.
Moreover, the expressions (\ref{eq21a}) and (\ref{eq21b}) do not depend on
the mass $m$, then the inertial term does not have significant influence on
the long-time behavior of the system. In Fig. 6 we verify the asymptotic
result of $x(t)$ obtained by Eqs. (\ref{eq20}) and (\ref{eq21a}), whereas in
Fig. 7 we verify the asymptotic result of $v(t)$ Eq. (\ref{eq20a}) by
comparing with Eq. (\ref{eq21b}). The curves converge to the same behavior
in the long-time limit. For small values of $m$ the oscillator does not move
across the final equilibrium position, whereas for large values of $m$ the
oscillator can move across the final equilibrium position. In other words,
the amplitude of oscillation increases with the increase of mass $m$. This
result is due to the inertial term which has important influence for the
initial movement of the system. We note that the solutions (\ref{eq20}) and (%
\ref{eq20a}) decay slowly due to the power-law asymptotic behavior. These
results are in contrast to those analyzed previously.

\section{Conclusion}

\bigskip

In this work we have investigated the motion of a particle governed by the
classical Newtonian equation (\ref{eq1}) under the influence of the
combination of local and nonlocal dissipative forces, linear external force
given by $U(x)=-kx$ and a constant load force $p$. This system can be used
to model a pressure control device. In particular, we have employed the
exponential and power-law functions for the dissipative memory kernel. Exact
and asymptotic solutions for the relaxation function $G(t)$, $x(t)$ and $%
v(t) $ have been obtained. The asymptotic results have permitted us to
obtain the final equilibrium position $x_{f}=p/k$. In the case of
exponential memory kernel the asymptotic results of $x(t)$ and $v(t)$ decay
exponentially similar to the system described by a local friction. For the
power-law memory kernel we have shown the asymptotic results of $x(t)$ and $%
v(t)$ which decay as a power-law and the leading terms are independent of
the parameter of local friction $\gamma _{\delta }$ and mass $m$; This means
that the long-time memory friction may suppress the presence of an
instantaneous friction in the long-time limit. We note that the system
designed in Fig. 1 can be made experimentally. Then we hope the model
described by Eq. (\ref{eq1}) may be used to investigate the dynamics of
membranes. If necessary other kinds of function for dissipative memory
kernel can also be used and the solutions can be obtained by Eq. (\ref{eq6}).

\bigskip

\textbf{Acknowledgment.} K.S. Fa \ acknowledges partial financial support
from \ the Conselho Nacional de Desenvolvimento Cient\'{\i}fico e Tecnol\'{o}%
gico (CNPq), \ Brazilian agency.

\newpage

\begin{center}
\textbf{Figure Captions}
\end{center}

\bigskip

\bigskip Fig. 1 - \ The elements of a pressure control system.

Fig. 2 - \ Behavior of $x(t)$ for different values of the mass $m$, in
arbitrary units. The parameters $k,$ $p$ and $\gamma _{\delta }$ have the
following values: $k=0.25$, $p=0.1$ and $\gamma _{\delta }=0.3.$ The dotted
line is obtained by Eq (\ref{eq10c}) with $m=0.09$.

Fig. 3 - Behavior of $v(t)$ for different values of the mass $m$, in
arbitrary units. The parameters $k,$ $p$ and $\gamma _{\delta }$ have the
following values: $k=0.25$, $p=0.1$ and $\gamma _{\delta }=0.3.$ The dotted
line is obtained by Eq (\ref{eq10d}) with $m=0.09$.

\bigskip

Fig. 4 - Behavior of $x(t)$ for different values of the mass $m$, in
arbitrary units. The parameters $\lambda $, $k,$ $p$, $\gamma _{\delta }$
and $\gamma _{\lambda }$ have the following values: $\lambda =0.1$, $k=0.25$%
, $p=0.1$, $\gamma _{\delta }=0.3$ and $\gamma _{\lambda }=0.13.$ The dotted
line is obtained by Eq (\ref{eq17a}).

\bigskip

Fig. 5 - Behavior of $v(t)$ for different values of the mass $m$, in
arbitrary units. The parameters $\lambda $, $k,$ $p$, $\gamma _{\delta }$
and $\gamma _{\lambda }$ have the following values: $\lambda =0.1$, $k=0.25$%
, $p=0.1$, $\gamma _{\delta }=0.3$ and $\gamma _{\lambda }=0.13.$ The dotted
line is obtained by Eq (\ref{eq17b}).

Fig. 6 - Behavior of $x(t)$ for different values of the mass $m$, in
arbitrary units. The parameters $\beta $, $k,$ $p$, $\gamma _{\delta }$ and $%
\gamma _{\beta }$ have the following values: $\beta =0.5$, $k=0.25$, $p=0.1$%
, $\gamma _{\delta }=0.3$ and $\gamma _{\beta }=0.13.$ The dotted line is
obtained by Eq (\ref{eq21a}).

Fig. 7 - Behavior of $v(t)$ for different values of the mass $m$, in
arbitrary units. The parameters $\beta $, $k,$ $p$, $\gamma _{\delta }$ and $%
\gamma _{\beta }$ have the following values: $\beta =0.5$, $k=0.25$, $p=0.1$%
, $\gamma _{\delta }=0.3$ and $\gamma _{\beta }=0.13.$ The dotted line is
obtained by Eq (\ref{eq21b}).


\begin{thebibliography}{99}
\bibitem{cox} I. Golding, E.C. Cox, Physical nature of bacterial cytoplasm, 
\textit{Phys. Rev. Lett.} \textbf{96} (2006) 0981021-4.

\bibitem{kou} S.C. Kou, X.S. Xie, Generalized langevin equation with
fractional Gaussian noise: Subdiffusion within a single protein molecule, 
\textit{Phys. Rev. Lett.} \textbf{93} (2004) 1806031-4.

\bibitem{min} W. Min, G. Luo, B.J. Cherayil, S.C. Kou, X.S. Xie, Observation
of a power-law memory kernel for fluctuations within a single protein
molecule, \textit{Phys. Rev. Lett.} \textbf{94} (2005) 1983021-4.

\bibitem{chaudhury} S. Chaudhury, S.C. Kou, B.J. Cherayil, Model of
fluorescence intermittency in single enzymes, \textit{J. Phys. Chem.} 
\textbf{111} (2007) 2377-2384.

\bibitem{bagley} R.L. Bagley, P.J. Torvik, A theoretical basis for the
application of fractional calculus to viscoelasticity, \textit{J. Rheology } 
\textbf{27} (1983) 201-210.

\bibitem{makris} R.L.N. Makris, M.C. Constantinou, Fractional-derivative
Maxwell model for viscous dampers, \textit{J. Structural Eng. } \textbf{117}
(1991) 2708-2724.

\bibitem{shiti1} Yu.A. Rossikhin, M.V. Shitikova, Application of fractional
operators to the analysis of damped vibrations of viscoelastic single-mass
systems, \textit{J. Sound and Vibration } \textbf{199}(1997) 567-586.

\bibitem{shiti2} Yu.A. Rossikhin, M.V. Shitikova, \ A new method for solving
dynamic problems of fractional derivative viscoelasticity, \textit{Int. J.
Eng. Science } \textbf{39} (2001) 149-176.

\bibitem{mainardi} F. Mainardi, \textit{Fractals and Fractional Calculus in
Continuum Mechanics, }Springer, Wien, 1997, pp. 291-348.

\bibitem{agrawal} O.P. Agrawal, Stochastic analysis of dynamic systems
containing fractional derivatives, \textit{J. Sound and Vibration } \textbf{%
247} (2001) 927-938.

\bibitem{adhikari} J. Sieber, D.J. Wagg, S. Adhikari, On the interaction of
exponential non-viscous damping with symmetric nonlinearities, \textit{J.
Sound and Vibration} \ 314 (2008) 1-11.

\bibitem{kreys} E. Kreyszig, Advanced engineering mathematics (fourth
edition), John Wiley \& Sons, New York, 1981, Chapter 5.

\bibitem{gorenflo} A. Carpinteri, F. Mainardi, Fractals and Fractional
Calculus in Continuum Mechanics, Springer, Wien, 1997, pp. 223-276.

\bibitem{erdel} A. Erdelyi, W. Magnus, F. Oberhettinger, F.G. Tricomi,
Higher Transcendental Functions, Vol. III, \ McGraw-Hill, USA, 1955.
\end{thebibliography}
\end{document}